\documentclass[twocolumn,showpacs,prl]{revtex4}

\usepackage{graphicx}
\usepackage{amssymb}

\begin{document}

\title{Scaling the Temperature-dependent Boson Peak of Vitreous Silica \\ 
with the high-frequency Bulk Modulus derived from Brillouin Scattering Data}

\author{B. Ruffl\'e$^1$}
\author{S. Ayrinhac$^1$}
\author{E. Courtens$^1$}
\author{R. Vacher$^1$}
\author{M. Foret$^1$}
\author{A. Wischnewski$^2$}
\author{U. Buchenau$^2$}
\affiliation{$^1$Laboratoire des Collo\"{\i}des, Verres et Nanomat\'eriaux, UMR 5587 CNRS \\
Universit\'e Montpellier II, F-34095 Montpellier Cedex 5, France \\
$^2$Institut f\"{u}r Fesk\"{o}rperforschung, Forschungszentrum J\"{u}lich, Postfach 1913, D-52425 J\"{u}lich, Germany}
\date{September 18, 2009}

\begin{abstract}
The position and strength of the boson peak in silica glass vary considerably with temperature $T$.
Such variations cannot be explained solely with changes in the Debye energy.
New Brillouin scattering measurements are presented which allow determining the $T$-dependence of unrelaxed acoustic velocities.
Using a velocity based on the bulk modulus, scaling exponents are found which agree with the soft-potential model.
The unrelaxed bulk modulus thus appears to be a good measure for the structural evolution of silica with $T$ and to set the energy scale for the soft potentials. 
\end{abstract}


\maketitle

The nature of collective vibrations in glasses and their relation to structural disorder are topics of active discussion and considerable interest.
The reduced density of vibrational states, $g ( \nu ) / \nu ^2 $, where $\nu = \omega / 2 \pi$ is the frequency, generally shows an excess over the Debye level $g _{\rm D}( \nu ) / \nu ^2 $ calculated from the acoustic velocities.
This excess, $I(\nu) \equiv (g - g_{\rm D})/\nu^2$, is called the boson peak (BP).
It is generally agreed that the BP must bear relation to the strong scattering of acoustic modes leading to the plateau in the temperature ($T$) dependence of the thermal conductivity, a feature universally observed in dielectric glasses \cite{Zel71}.
Two broad categories of processes are mainly invoked to explain these anomalies as summarized {\em e.g.} in \cite{Ruf08,Bal09}: 1) the interaction of acoustic modes with structural or elastic disorder, and 2) the presence of additional vibrations and their resonant coupling to acoustic waves, as described {\em e.g.} by the soft-potential model (SPM) \cite{Par94} or related developments \cite{Par07}.
Several authors recently attempted scaling BP data in terms of the Debye density of states, {\em e.g.} \cite{Mon06a,Mon06b,Nis07,Cap09,Bal09}.
They advocated that such a scaling supports the view that BP modes are strictly acoustic.
We examine here the case of silica, a prototypical glass of high technical interest which exhibits the strongest known BP excess \cite{Sok97a}.
We find that scaling with the Debye velocity is inappropriate, while scaling in terms of the bulk modulus leads both to a satisfactory master curve and to exponents that are compatible with the SPM.
We propose that an appropriately determined bulk modulus is a good measure for the structural evolution of silica with $T$.
Further, the relation to the SPM implies that the second category of models is here the relevant one.

Silica is a good candidate for a meaningful scaling of the BP in function of $T$, since both the BP position, $\nu_{\rm BP}$, and strength, $I_{\rm BP} \equiv I(\nu_{\rm BP})$, vary significantly with $T$ \cite{Wis98}.
The glass exhibits anomalous thermomechanical properties that are typical for tetrahedral networks \cite{Kra68}.
Among them, the elastic moduli decrease under pressure \cite{Vuk72} and harden with increasing $T$ \cite{Pol02}.
Simulations of silica indicate a progressive and reversible polyamorphic transformation related to the reorientations of the --Si--O--Si-- bonds forming ring structures, this without bond breaking or reconstruction \cite{Hua04}.
The BP evolution presumably relates to that transformation.
A good measure for the degree of transformation might be a {\em suitably defined} elastic modulus.
One should recall that the elastic properties of glasses at ultrasonic frequencies are affected by thermally activated relaxations (TAR) of structural defects, as known for over half-a-century  \cite{And55}.
Furthermore, the anharmonic coupling of sound with the thermal bath depresses the sound velocities with increasing $T$, also observed long ago \cite{Cla78}.
A {\em suitably defined} structure-dependent modulus should not include these viscoelastic effects.
The particular case of silica was recently revisited on the basis of available and new measurements of sound velocity and attenuation covering a very broad range of $\nu$ and $T$ \cite{Vac05}.
The relaxations can be described by double-well potentials with a distribution of barriers and asymmetries \cite{Gil81}. The appropriate distribution and the effect of anharmonicity were determined in \cite{Vac05}. 
This allowed extracting an unrelaxed or {\em bare} velocity for the longitudinal acoustic mode, $v_{\infty}^{\rm LA}$, which was found to increase considerably with $T$ \cite{Vac05}.
As shown below, the same is now observed on the transverse mode, $v_{\infty}^{\rm TA}$.
The bare bulk modulus, $K_\infty$, also increases strongly with $T$.
Changes in the corresponding ``velocity'', $v_{\infty}^{\rm K} \propto \sqrt{K_\infty}$, could provide a measure for the progress in the polyamorphic transformation with increasing $T$.
In this Letter, we show that indeed the BP of silica successfully scales unto a single master curve with the use of $v_{\infty}^{\rm K} (T)$, this with exponents that are non-trivial.
Our results indicate that the bare modulus is a good measure for the structural evolution in function of $T$, and suggest that the latter affects the strength and position of the BP.

The symbols in Figure 1 present Brillouin-scattering results that are new for LA waves at elevated $T$ as well as for TA waves over the entire range.
The data were obtained on a high-quality silica sample of low OH concentration ($\leq 100$ ppm) using the high-resolution tandem interferometer described in \cite{Rat05}.
The Brillouin shifts $\Omega$ and half-widths $\Gamma$ are measured near backscattering (LA) or at $90^{\circ}$ (TA).
Care is taken to eliminate the spectral broadening due to the finite aperture, which is an important correction at $90^{\circ}$.
The shifts are converted to sound velocities using the known $T$-dependence of the refractive index \cite{Ayr08}.
\begin{figure}
\includegraphics[width=8.5cm]{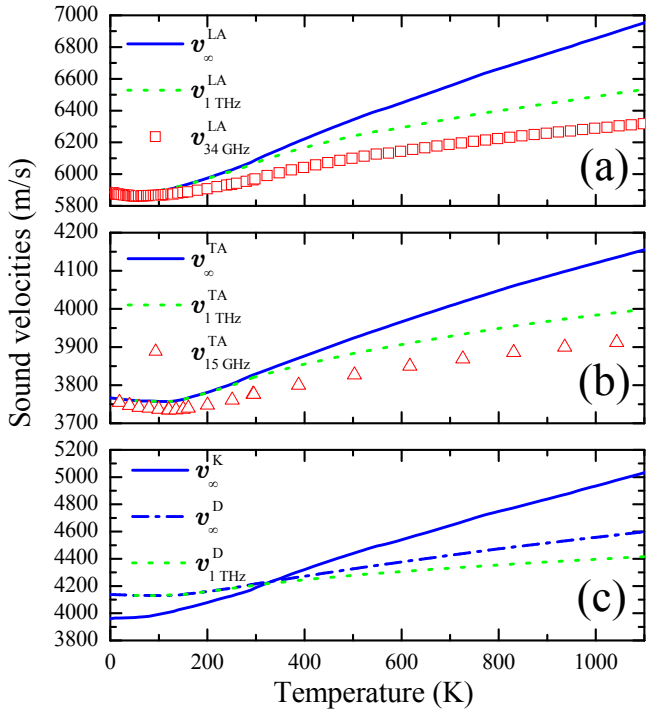}
\caption{The $T$-dependence of the sound velocities measured in silica with Brillouin scattering (points) and renormalized to 1 THz (dashed lines) and to infinite frequency (solid lines): (a) the LA mode; (b) the TA mode; (c) the calculated Debye velocities $v^{\rm D}$ at two frequencies compared to the bare bulk-modulus ``velocity'' $v_{\infty}^{\rm K}$.} 
\end{figure}
Following \cite{Vac05}, the internal friction $Q^{-1} = 2 \Gamma / \Omega$ allows calculating the contributions of TAR and anharmonicity to the velocities, $\delta v _{\rm TAR}$ and $\delta v _{\rm ANH}$, respectively \cite{foot1}.
Correcting the data points in Fig. 1 for these velocity shifts, the solid lines representing the bare velocities for both LA and TA modes are obtained.
Near and above room $T$, it is the anharmonic term $\delta v _{\rm ANH} = - v Q^{-1}/\Omega \tau _{\rm th}$ which dominates by far the velocity corrections needed to obtain the bare values \cite{Vac05}.
The principal source of uncertainty is in the mean thermal relaxation time $\tau _{\rm th}$.
>From \cite{Vac05,Ayr08} we estimate that the uncertainty in $\ln \tau _{\rm th}$ is at most $\pm 0.1$, which leads to the same uncertainty on $\delta v _{\rm ANH}/v$.
The dashed lines show the velocities calculated at the intermediate frequency of 1 THz corresponding to the approximate position of the BP maximum, $\nu _ {\rm BP}$.
The $T$-dependence of the bare velocities, $v_{\infty}^{\rm LA}$ and $v_{\infty}^{\rm TA}$,
is considerably stronger than observed at Brillouin-scattering frequencies.
It should be remarked that in the absence of structural changes with $T$ and with negligible density changes, the bare velocities should be {\em independent} of $T$.
The observed dependence is thus a signature of the progressive polyamorphic transformation \cite{Hua04}.
As already mentioned in \cite{Vac05,Rat05}, the velocities $v_\infty$ might be hard to directly observe.
The reason is the interaction with the BP, as described {\em e.g.} in \cite{Ruf08,Dev08}.
However, at constant density these velocities directly relate to microscopic elastic stiffnesses.
Based on $v_{\infty}^{\rm LA}$ and $v_{\infty}^{\rm TA}$ one can construct other quantities.
If the interest is in the density of acoustic modes, one considers the unrelaxed Debye velocity $v_{\infty}^{\rm D}$ given by $3/(v_{\infty}^{\rm D})^3 = 1/(v_{\infty}^{\rm LA})^3 + 2/(v_{\infty}^{\rm TA})^3$.
If instead the interest is in the average rigidity of the structure at short distances, one can consider a ``velocity'' $v_{\infty}^{\rm K}$ given by $(v_{\infty}^{\rm K})^2 = (v_{\infty}^{\rm LA})^2 - {\frac{4}{3}}(v_{\infty}^{\rm TA})^2$ since the bulk modulus $K$ relates to the elastic constants by $K = C_{11} - {\frac{4}{3}}C_{44}$.
The very different $T$-dependence of $v_{\infty}^{\rm D}$ and $v_{\infty}^{\rm K}$is emphasized in Fig. 1c.
We now explore the relation between the bare velocities and the BP position and strength.
\begin{figure}
\includegraphics[width=8.5cm]{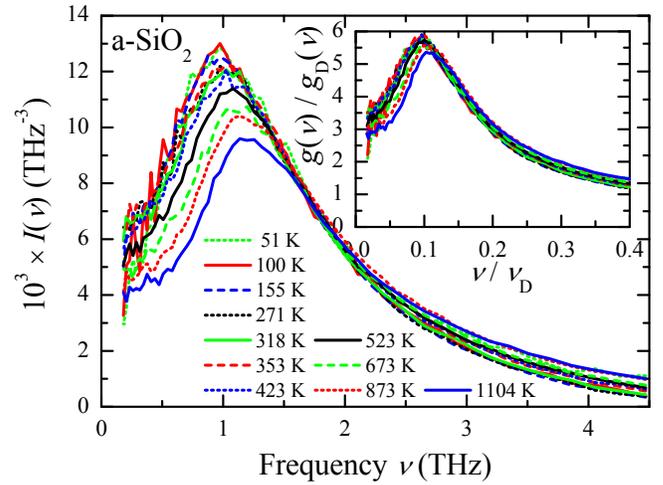}
\caption{The neutron scattering BP of silica at 11 different temperatures. The Debye level at 1 THz calculated from the velocity in Fig. 1c is already subtracted. The inset illustrates a scaling of the entire $g(\nu)$ using the Debye frequency at 1 THz, the BP frequency.} 
\end{figure}

Figure 2 shows measurements of the excess density of vibrational states of silica in the BP region. 
The data were obtained with neutron scattering as described in \cite{Wis98}.
For selection rule reasons, it is most important to use here neutron data rather than Raman scattering ones as available {\em e.g.} in \cite{Fon99}.
Indeed, ${\rm SiO}_4$-libration modes that are inactive in Raman scattering are important to the BP \cite{Buc86}, as also confirmed in a hyper-Raman study \cite{Heh00}.
The ordinate of Fig. 2 shows the excess $I(\nu)$, obtained by subtracting from the various curves the Debye level, $g _{\rm D}( \nu ) / \nu ^2 $.
For this calculation we used a constant Debye wave vector $k_{\rm D} = 1.576 \times 10^{10}$ m$^{-1}$ as the variation of the atomic density with $T$ is comparatively negligible.
The Debye level is then $3/\nu_{\rm D}^3$, with $\nu_{\rm D} = v_{\rm D} k_{\rm D} / 2\pi$, where $v_{\rm D}$ is taken from Fig. 1c at 1 THz.
One notices that $I(\nu)$ does not approach 0 at $\nu = 0$.
The reason lies in quasi-elastic scattering (QES), a low frequency strongly anharmonic excess that scatters in addition to the harmonic BP.
The two components can be separated based on the anharmonicity, revealing that QES ``decreases with increasing $\nu$ and is undetectable above 600 GHz at room $T$ and below'' \cite{Buc88}.
The QES contribution also goes through a maximum below room $T$, while it decreases and does not show any additional broadening at elevated $T$ \cite{Sok97b,Fon05}.
In view of this, and as it would be difficult subtracting QES from the data, we rather leave it but do not insist that scaling applies below 0.65 THz.
The inset of Fig. 2 zooms on the BP region, using a scaled abscissa, $\nu/\nu_{\rm D}$, and a scaled ordinate, $g/g_{\rm D}$.
This Debye scaling does not lead to a satisfactory master curve, not so much because of a poor scaling of the intensities, but mainly because the BP positions do not superpose.
This situation is not significantly improved if one used $v_{\infty}^{\rm D}$ in place of $v_{\rm 1 \, THz}^{\rm D}$, as will become clear below.

We now obtain from Fig. 2 the $T$-dependence of $\nu_{\rm BP}$ and $I_{\rm BP}$ that are shown in Fig. 3.
To this effect, the successive curves are scaled to the first one taken as reference.
The data at 51 K are indeed least affected by QES.
Specifically, the curve at $T$ is scaled by replacing $\nu$ by $\nu/x$ and $I$ by $I/y$.
Its difference with the 51 K curve is then minimized by a least-square procedure over the range from $\nu_{\rm BP} - 0.35$ THz to $\nu_{\rm BP} + 1$ THz.
The BP parameters at $T$ are then $\nu_{\rm BP}(T) = x \: \nu_{\rm BP}(51 \; \rm K)$ and $I_{\rm BP}(T) = y \: I_{\rm BP}(51 \; \rm K)$.
\begin{figure}
\includegraphics[width=8.5cm]{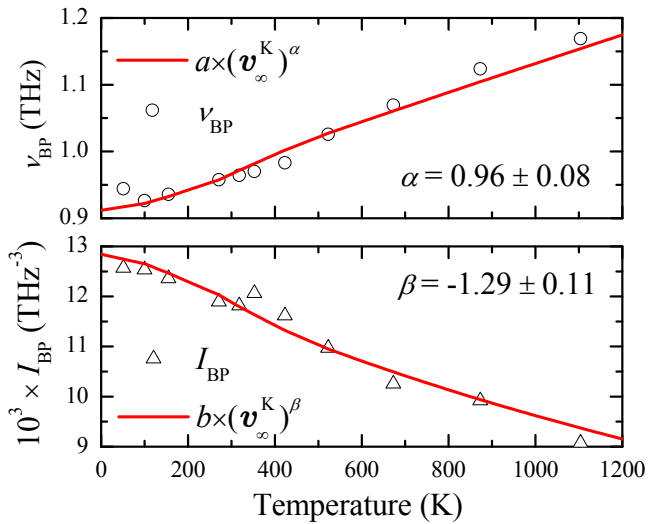}
\caption{The experimentally determined BP positions and intensities in function of $T$, adjusted to powers of $v_{\infty}^{\rm K}$.} 
\end{figure}
To obtain the values on an absolute scale, it remains to estimate $\nu_{\rm BP}(51 \: \rm K)$ and $I_{\rm BP}(51 \: \rm K)$.
This is done by fitting the 51 K data to a log-normal, $I_{\rm BP} \, {\rm exp} [-({\rm log} \, \nu /\nu_{\rm BP})^2 / 2 \sigma ^2]$.
Although this is somewhat {\em ad hoc}, it is of no real importance since absolute values do not affect the scaling exponents to be determined below.
As observed in Fig. 3, it is remarkable that $\nu_{\rm BP}$ increases by as much as 24\% and that $I_{\rm BP}$ decreases
by 39\% over this $T$ range.
Over the same range of $T$, $v_{\rm 1 \, THz}^{\rm D}$ only increases by 7\%, while $v_{\infty}^{\rm D}$ increases by 11\%.
Here, a Debye scaling implies that $\nu_{\rm BP} \propto \nu_{\rm D} \propto v^{\rm D}$ 
and that $I_{\rm BP} \propto \nu_{\rm D}^{-3} \propto (v^{\rm D})^{-3}$.
While the latter happens to be approximately verified with $v_{\infty}^{\rm D}$, the former cannot.
This shows that checking the validity of the Debye scaling can be a delicate matter to which we return in the final discussion.
At any rate this scaling does not work in silica.
This is not so surprising.
Indeed, there is now ample evidence that the BP of silica does {\em not} derive its strength from acoustic modes, as already known from neutron scattering \cite{Buc86} and hyper-Raman \cite{Heh00} results.

\begin{figure}
\includegraphics[width=8.5cm]{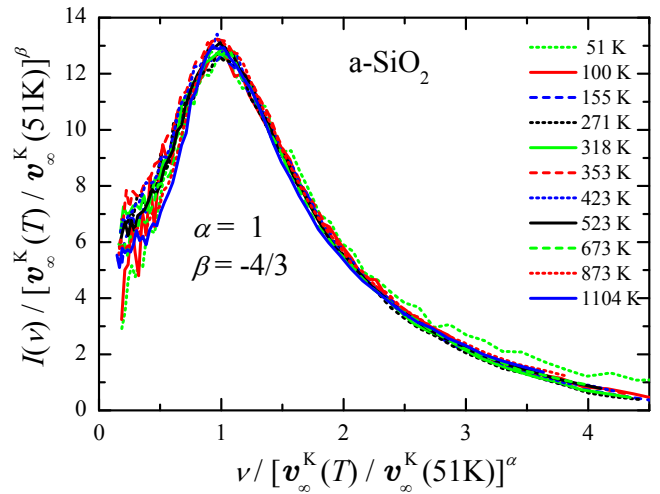}
\caption{The BP data of Fig. 2 scaled with exponents $\alpha = 1$ and $\beta = -4/3$.} 
\end{figure}
The data points of Fig. 3 can be adjusted to a bare velocity $v_\infty$ with 
$$\nu_{\rm BP}(T) = a [v_\infty (T)]^\alpha \;\;\; , \;\;\; I_{\rm BP}(T) = b [v_\infty (T)]^\beta \; . \eqno{(1)}$$
We remark that if this works for one particular type of $v_\infty$, it will work for all.
Indeed, over the restricted ranges of interest here, we observe the approximate relations $v_{\infty}^{\rm K} \sim (v_{\infty}^{\rm LA})^{1.43} \sim (v_{\infty}^{\rm D})^{2.13} \sim (v_{\infty}^{\rm TA})^{2.27}$.
We note that $v_{\infty}^{\rm K}$ increases by 25\% over the range of $T$, rather similar to the increase in $\nu_{\rm BP}$.
It seems thus appropriate to first try $v_{\infty}^{\rm K}$ in (1).
This gives the solid lines traced in Fig. 3 with the exponents shown there.
The scatter in the experimental points does not result from the scaling procedure described in the previous paragraph, but rather from the neutron data themselves.
The uncertainty in $v_\infty^{\rm K}$ related to $\tau _{\rm th}$ leads to variations in both $\alpha $ and $\beta $ that are about half the error bars given in Fig. 3.
The exponents are nearly $\alpha = 1$ and $\beta = -4/3$, well within these error bars.
Using the latter values, the entire data scales as shown in Fig. 4.
Except for the region below $\sim 0.75$ THz which is affected by QES, the scaling is obviously very satisfactory.
It is now of interest to consider the meaning of these exponents within the SPM.

The progressive polyamorphic transformation of silica occurs without any change in the network connections \cite{Hua04}.
Hence, the number of defects producing quasi-local vibrations should not change.
It is the environment of the soft harmonic oscillators which is modified.
The latter are characterized by an energy ${\cal E}_0 = M v^2$ \cite{Par94}, where $M$ is the mean atomic mass.
The unspecified velocity $v$ entering ${\cal E}_0$ is certainly an unrelaxed $v_{\infty}$.
Assuming that only ${\cal E}_0$ changes with $T$, and using Eqs. (1.5) of \cite{Par94}, one obtains
$$\eta_{\rm L} \, \propto \, v_ \infty ^ {-2/3} \;\;\; , \;\;\; W \, \propto \, v_ \infty ^{\; 2/3} \;\; . \eqno{(2)}$$
Here, $\eta_{\rm L}$ is the small parameter that scales the kinetic energy of the soft potential Hamiltonian, and $W$ is the crossover energy between vibrational and tunneling states.
The BP intensity is fully determined by the strength of its low frequency ($\nu \ll \nu_{\rm BP}$) onset.
This is seen by comparing Eqs. (5.12) and (5.18) of \cite{Par07}.
One can thus use a well-known expression for the onset, which is that $I(\nu) \propto \nu^2 / W^5$ \cite{Ram93}, up to $\nu = \nu_{\rm BP}$ to derive the scaling. 
This gives
$$\nu_{\rm BP}^2 \, / \, I_{\rm BP} \, \propto \,  W^5 \;\; . \eqno{(3)}$$
Introducing (1) and (2) in (3), one obtains
$$ 2 \alpha \, - \, \beta \, = \, 10/3 \;\; . \eqno{(4)}$$
This precisely agrees with the values $\alpha = 1$ and $\beta = -4/3$ found above using $v_{\infty}^{\rm K}$ for scaling.
This suggests that the bulk modulus gives in the present case a sufficiently appropriate measure for the average interactions of the soft-potentials with their environment.
${\cal E}_0$ being controlled by an inverse compressibility, these interactions seem to be mostly hydrodynamic-like on the average. 

Compared to silica in function of $T$, in the silicates that were investigated for scaling, the relative range of $\nu_{\rm BP}$ is smaller.
It is about 6\% for the three curves that scale in  \cite{Mon06b}, less in \cite{Mon06a}, and nil in \cite{Bal09}, while it is 24\% presently.
That makes checking for the validity of a Debye scaling all the more demanding.
It would require a stringent analysis both of the peak positions and of the intensity which is in {\em excess} over the Debye level.
In particular in \cite{Bal09}, there is no change in $\nu_{\rm D}$ and thus no possibility to check the scaling law.
By comparison, there exists one report of a failure of the Debye scaling tested on a polymer under pressure \cite{Nis07}.
A similar conclusion was anticipated in \cite{Hon08}.
On the other hand there is one report of a successful Debye scaling of Raman scattering data on a reactive mixture during polymerization \cite{Cap09}.
However, this is a complicated physico-chemical situation so that the significance of the result is momentarily not understood.
Summarizing, it would be hard concluding from available scaling evidence that the origin of boson peaks in glasses is necessarily acoustic.

Our results show that for silica in function of $T$ a Debye scaling of the large excursions in $\nu_{\rm BP}(T)$ and $I_{\rm BP}(T)$ is not possible.
A scaling can be performed in terms of unrelaxed velocities $v_\infty$.
The exponents that are found using a bare velocity based on the bulk modulus, $v_{\infty}^{\rm K}$, are remarkably compatible with the existence of quasi-local vibrations described by the soft-potential model.
It thus seems that the unrelaxed bulk modulus provides a good measure for the $T$-dependent polyamorphic transformation of silica and that it plays a key role in setting the scale for the the soft potentials.


\begin{references}

\bibitem{Zel71}R.C. Zeller and R.O. Pohl, Phys. Rev. B {\bf 4}, 2029 (1971).

\bibitem{Ruf08}B. Ruffl\'{e}, D.A. Parshin, E. Courtens, and R. Vacher, Phys. Rev. Lett. {\bf 100}, 015501 (2008).

\bibitem{Bal09}G. Baldi {\em et al.}, Phys. Rev. Lett. {\bf 102}, 195502 (2009).

\bibitem{Par94}D.A. Parshin, Fiz. Tverd. Tela (Leningrad) {\bf 36}, 1809 (1994) [Sov. Phys. Solid State {\bf 36}, 991 (1994)].

\bibitem{Par07}D.A. Parshin, H.R. Schober, and V.L. Gurevich, Phys. Rev. B {\bf 76}, 064206 (2007).

\bibitem{Mon06a}A. Monaco {\em et al.}, Phys. Rev. Lett. {\bf 96}, 205502 (2006).

\bibitem{Mon06b}A. Monaco {\em et al.}, Phys. Rev. Lett. {\bf 97}, 135501 (2006).

\bibitem{Nis07}K. Niss {\em et al.}, Phys. Rev. Lett. {\bf 99}, 055502 (2007).

\bibitem{Cap09}S. Caponi {\em et al.}, Phys. Rev. Lett. {\bf 102}, 027402 (2009).

\bibitem{Sok97a}A.P. Sokolov {\em et al.}, Phys. Rev. Lett. {\bf 78}, 2405 (1997).

\bibitem{Wis98}A. Wischnewski, U. Buchenau, A.J. Dianoux, W.A. Kamitakahara, and J.L. Zarestky, Phys. Rev. B {\bf 57}, 2663 (1998).

\bibitem{Kra68}J.T. Krause and C.R. Kurkjian, J. Am. Ceram. Soc. {\bf51}, 226 (1968).

\bibitem{Vuk72}M.R. Vukevich, J. Non-Cryst. Solids, {\bf 11}, 25 (1972).

\bibitem{Pol02}A. Polian {\em et al.}, Europhys. Lett. {\bf57}, 375 (2002).

\bibitem{Hua04}Liping Huang and J. Kieffer, Phys. Rev. B {\bf 69}, 224203 (2004).

\bibitem{And55}O.L. Anderson and H.E. B\"{o}mmel, J. Am. Ceram. Soc. {\bf 38}, 125 (1955).

\bibitem{Cla78}T.N. Claytor and R.J. Sladek, Phys. Rev. B {\bf 18}, 5842 (1978).

\bibitem{Vac05}R. Vacher, E. Courtens, and M. Foret, Phys. Rev. B {\bf 72}, 214205 (2005).

\bibitem{Gil81}K.S. Gilroy and W.A. Phillips, Philos. Mag. {\bf 43}, 735 (1981).

\bibitem{Rat05}E. Rat {\em et al.}, Phys. Rev. B {\bf 72}, 214204 (2005).

\bibitem{Ayr08}S. Ayrinhac, Doctoral thesis, Univ. of Montpellier 2, Nov. 2008 (unpublished).

\bibitem{foot1}With the new $\Gamma$ values of the present measurements, the constant ${\cal C}$ scaling the TAR contribution in \cite{Vac05} now equals $1.9 \times 10^{-3}$ \cite{Dev08}, the same for LA and TA modes. 

\bibitem{Dev08}A. Devos {\em et al.}, Phys. Rev. B {\bf 77}, 100201(R) (2008).

\bibitem{Fon99}A. Fontana {\em et al.}, Europhys. Lett. {\bf 47}, 56 (1999).

\bibitem{Buc86}U. Buchenau {\em et al.}, Phys. Rev. B {\bf 34}, 5665 (1986). 

\bibitem{Heh00}B. Hehlen {\em et al.}, Phys. Rev. Lett. {\bf 84}, 5355 (2000).

\bibitem{Buc88}U. Buchenau {\em et al.}, Phys. Rev. Lett. {\bf 60}, 1318 (1988).

\bibitem{Sok97b}A.P. Sokolov {\em et al.}, Europhys. Lett. {\bf 38}, 49 (1997).

\bibitem{Fon05}A. Fontana {\em et al.}, J. Non-Cryst. Solids {\bf 351}, 1928 (2005).

\bibitem{Ram93}M.A. Ramos {\em et al.}, phys. sta. sol. (a) {\bf 135}, 477 (1993).

\bibitem{Hon08}L. Hong {\em et al.}, Phys. Rev. B {\bf 78}, 134201 (2008).



\end{references}
\end{document}